\documentclass[11pt]{article}

\usepackage{latexsym,amssymb}

\newtheorem{theorem}{Theorem}
\newtheorem{definition}[theorem]{Definition}

\newtheorem{corollary}[theorem]{Corollary}
\newtheorem{example}[theorem]{Example}

\newtheorem{fact}{Fact}

\def\C{{\mathbb{C}}}

\newcommand{\cB}[0]{{\mathcal B}}

\newcommand{\onemat}[0]{{\mathbf 1}}

\newcommand{\proof}[1]{\medskip \noindent {\em Proof.} #1 \hfill $\Box$\vspace{0.5cm}}

\newcommand{\ket}[1]{|#1\rangle}

\newcommand{\scalar}[2]{\langle #1|#2\rangle}

\newcommand{\m}[0]{{\bf m}}

\begin{document}

\title{\Large \textbf{New Construction of Mutually Unbiased Bases\\ in
Square Dimensions}}

\author{Pawel Wocjan and Thomas Beth\\
Institut f\"ur Algorithmen und Kognitive Systeme\\
Universit\"at Karlsruhe, 76\,128 Karlsruhe, Germany\\
email: \texttt{wocjan\symbol{64}ira.uka.de}
}

\date{July 12, 2004}

\maketitle

\begin{abstract}
We show that $k=w+2$ mutually unbiased bases can be constructed in any
square dimension $d=s^2$ provided that there are $w$ mutually
orthogonal Latin squares of order $s$. The construction combines the
design-theoretic objects $(k,s)$-{\it nets} (which can be constructed
from $w$ mutually orthogonal Latin squares of order $s$ and vice
versa) and generalized Hadamard matrices of size $s$. Using known
lower bounds on the asymptotic growth of the number of mutually
orthogonal Latin squares (based on number theoretic sieving
techniques), we obtain that the number of mutually unbiased bases in
dimensions $d=s^2$ is greater than $s^{1/14.8}$ for all $s$ but
finitely many exceptions. Furthermore, our construction gives more
mutually orthogonal bases in many non-prime-power dimensions than the
construction that reduces the problem to prime power dimensions.
\end{abstract}

%
%

\section{Introduction}
Two orthonormal bases ${\cal B}$ and ${\cal B}'$ of the Hilbert space
$\C^d$ are called {\it mutually unbiased} if and only if
\[
|\scalar{\phi}{\psi}|^2=1/d
\]
for all $\ket{\phi}\in {\cal B}$ and all $\ket{\psi}\in {\cal
B}'$. The concept of {\it mutually unbiased bases} (MUBs) plays an
important role in quantum information theory. For example, some
quantum cryptographic protocols rely on the fact that no information
can be obtained when a quantum system which is initialized in a basis
state from ${\cal B}'$ is measured with respect to the basis ${\cal
B}$. The protocols in \cite{BB:84,Bruss:98,BT:00} exploit this
observation in order to distribute secrete keys over a public channel
in an information-theoretic secure way.

It is known that the cardinality of any collection of mutually
unbiased bases of $\C^d$ cannot exceed $d+1$ (see
\cite{BBRV:02,DGS:75,Hoggar:82,KL:78,WF:89}). Sets attaining this
bound are extremely interesting because they allow quantum state
tomography with projective measurements consisting of a minimal number
of operators \cite{Ivanovic:81}. Furthermore, for some of these cases
it is possible to define a discrete version of the Wigner function for
states on the Hilbert space $\C^d$ (see e.g.\ \cite{GHW:04} for an
overview on this subject). An enjoyable application of mutually
unbiased bases is the Mean King's problem
\cite{VA:87,EA:01,Aravind:03}.

Let $N_{MUB}(d)$ denote the maximum cardinality of any set containing
pairwise mutually unbiased bases of $\C^d$. We have $N_{MUB}(d)\ge 3$
for any dimension $d$ (see \cite{BBRV:02}). It is known that
$N_{MUB}(d)=d+1$ holds when $d$ is a prime power
\cite{Ivanovic:81,WF:89,Zauner:99,BBRV:02,KR:03}.

An open problem is to determine $N_{MUB}(d)$ when the dimension is not
a prime power. The case $d=6$ is studied in \cite{Grassl:04}. An
elementary lower bound on $N_{MUB}(d)$ for arbitrary $d$ is derived in
\cite[Proposition 2.20]{Zauner:99} and \cite{KR:03}. Let
$d=p_1^{e_1}\cdots p_r^{e_r}$ be a factorization of $d$ into distinct
prime powers $p_i^{e_i}$. Then
\begin{equation}\label{eq:backPrimePower}
N_{MUB}(d)\ge 
\min\{ 
N_{MUB}(p_1^{e_1}), 
N_{MUB}(p_2^{e_2}),\ldots,
N_{MUB}(p_r^{e_r})\}\,.
\end{equation}
This is proved as follows. Let $m:=\min_i N_{MUB}(p_i^{e_i})$. Choose
$m$ mutually unbiased bases ${\cal B}_1^{(i)}\ldots,{\cal B}_m^{(i)}$
of $\C^{p_i^{e_i}}$, for all $i$ in the range $1\le i \le r$. Then
\[
\{ 
{\cal B}_k^{(1)}\otimes \cdots\otimes {\cal B}_k^{(r)}\,\, | \,\,
k=1,\ldots,m
\}
\]
is a set of $m$ mutually unbiased bases of $\C^d$. This is easily
memorable by $N_{MUB}(mn)\ge \min\{N_{MUB}(n),N_{MUB}(m)\}$ for all
$m,n\ge 2$. In the following we will refer to this construction as the
``reduce to prime powers'' construction.

We present a construction that yields more MUBs than the ``reduce to
prime powers'' construction in many {\it non-prime-power
dimensions}. However, our construction always produces less MUBs if
$d^2$ is a prime power. More precisely, if $d=p^{2e}$ for some prime
number and $e>1$ then our construction gives only $p^e+1$ MUBs
compared to the optimal value $N_{MUB}(p^{2e})=p^{2e}+1$.

Our construction combines the design-theoretic objects nets or, more
precisely, their description by incidence vectors, and generalized
Hadamard matrices. These concepts are briefly introduced in
Section~2. The construction is presented in Section~3. In Section~4 we
discuss briefly the equivalence of nets to other design-theoretic
objects. We relate mutually orthogonal Latin squares and nets in more
detail. This shows that there are incidence vectors with good
parameters. In Section~5 we examine the cases where our construction
yields more mutually orthogonal bases.

\section{The ingredients}

\subsection{Incidence vectors from nets}\label{sub:incidence}
Our construction of mutually unbiased bases makes use of incidence
vectors satisfying certain conditions. The idea to use them is based
on the observation that mutually unbiased bases satisfy ``similar''
conditions. Let
\begin{eqnarray*}
\cB_1 & := & \{\ket{\psi_{11}},\ket{\psi_{12}},\ldots,\ket{\psi_{1d}}\}\,, \\
\cB_2 & := & \{\ket{\psi_{21}},\ket{\psi_{22}},\ldots,\ket{\psi_{2d}}\}\,, \\
\vdots && \\
\cB_k & := & \{\ket{\psi_{k1}},\ket{\psi_{k2}},\ldots,\ket{\psi_{kd}}\} 
\end{eqnarray*}
be a collection of $k$ mutually unbiased bases in $\C^d$.  For such a
collection we have
\begin{equation}\label{eq:orthogonal}
| \langle \psi_{bi} | \psi_{bj} \rangle |^2 = \delta_{ij} 
\end{equation}
for all $1\le b\le k$ and all $1\le i,j\le d$ and
\begin{equation}\label{eq:unbiased}
| \langle {\psi}_{bi} | \psi_{cj} \rangle |^2 = \frac{1}{d}
\end{equation}
for all $1\le b < c \le k$ and all $1\le i\le j\le d$. 

Now we consider a collection of incidence vectors that satisfy
``similar'' conditions. We say that a (column) vector
$\m:=(\m[1],\ldots,\m[d])^T$ of size $d$ is an {\it incidence vector}
if its entries take only the values $0$ and $1$. The {\it Hamming
weight} of $\m$ is the number of $1$'s. The support ${\rm supp}(\m)$
of the incidence vector $\m$ (of Hamming weight $s$) is the set of
indices $j_1,\ldots,j_s$ such that the corresponding entries
$\m[j_1],\ldots,\m[j_s]$ of $\m$ are all $1$. We always sort the
indices, i.e., $j_1<\ldots<j_s$.

Now we introduce the notion of nets, i.e., collections of incidence
vectors satisfying special properties.

\begin{definition}[Net]${}$\\
Let
$\{\m_{11},\ldots,\m_{1s},\m_{21},\ldots,\m_{2s},\ldots,\m_{k1},\ldots,\m_{ks}\}$
be a collection of $ks$ incidence vectors of size $d=s^2$ that are
partitioned into $k$ {\it blocks} where each block contains $s$
incidence vectors. The incidence vectors are denoted by $\m_{b i}$,
where $b=1,\ldots,k$ identifies the {\it block} and $i=1,\ldots,s$ the
vector within a {\it block}. If the incidence vectors satisfy the
following conditions we say that they form a $(k,s)$-net.
\begin{enumerate}
\item The supports of all vectors within one block are disjoint, i.e.,
\begin{equation}\label{eq:disjoint}
\m_{bi}^T\, \m_{bj} = 0 
\end{equation}
for all $1\le b \le k$ and all $1\le i\neq j\le s$.
\item The intersection of any incidence vectors from two different
blocks contains exactly one element, i.e.,
\begin{equation}\label{eq:one}
\m_{bi}^T\, \m_{cj}=1
\end{equation}
for all $1\le b\neq c\le s$ and all $1\le i,j\le s$.
\end{enumerate}
\end{definition}
Note that our definition of $(k,s)$-nets is in accordance with the
usual definition of nets in design theory \cite[page 172]{CD:96}. The
incidence vectors are just the characteristic functions of the subsets
used there.

\begin{example}[Incidence vectors]\label{ex:incidence}${}$\\
For $d=s^2$ with $s=2$ we have $k=3$ blocks. The incidence vectors are
defined in the table below for a $(3,2)$-net:
\[
\begin{array}{c|c||c|c||c|c}
\m_{11} & \m_{12} & \m_{21} & \m_{22} &
\m_{31} & \m_{32} \\
\hline 
1 & 0 & 1 & 0 & 1 & 0 \\
1 & 0 & 0 & 1 & 0 & 1 \\
0 & 1 & 1 & 0 & 0 & 1 \\
0 & 1 & 0 & 1 & 1 & 0 \\ \hline 
\end{array}
\]
\end{example}
In Section~\ref{sec:equivalence} we describe briefly that nets are
equivalent to other combinatorial objects like {\it transversal
designs}, {\it mutually orthogonal Latin squares}, and {\it orthogonal
arrays}. Due to the correspondence of equations (\ref{eq:orthogonal})
$\leftrightarrow$ (\ref{eq:disjoint}) and (\ref{eq:unbiased})
$\leftrightarrow$ (\ref{eq:one}) one could consider mutually unbiased
bases as the quantum generalization of nets. Indeed, this and other
quantum generalizations of many design-theoretic objects are
considered in \cite{Zauner:99}.

\subsection{Generalized Hadamard matrices}
For our construction we need to introduce generalized Hadamard
matrices. We call an $s\times s$-matrix $H$ with entries from $\C$ a
{\it generalized Hadamard matrix} if all its entries have modulus one
and $H$ satisfies $H H^\dagger=s \onemat_s$. Usual Hadamard matrices
have entries $\pm 1$. A necessary condition for the existence of
(usual) Hadamard matrix is that $s=1,2$ or $s$ is a multiple of $4$
(see \cite{HSS:99}).

Note that generalized Hadamard matrices exist for any dimension
$s$. Just take the (not normalized) Fourier matrix $DFT_s$ whose
entries $(k,l)$ are given by the $s$-th roots of unity
\[
DFT^{(s)}_{k,l}:=\omega^{k l}\,
\]
where $\omega:=e^{2 i\pi/s}$ and $k,l=0,\ldots,s-1$. Other examples of
generalized Hadamard matrices are character tables of finite abelian
groups \cite{JL:2001}. This follows from the orthogonality relations
of characters.

\section{Mixing the ingredients: our construction} 
Now we have all ingredients we need for our construction. We just need
the following simple definition. Let $\m\in\{0,1\}^d$ be an incidence
vector of Hamming weight $s$ and ${\bf h}\in\C^s$ an arbitrary column
vector. Then we define {\it the embedding of ${\bf h}$ into $\C^d$
controlled by $\m$}, denoted by ${\bf h}\uparrow\m$, to be the
following vector in $\C^d$
\begin{equation}
{\bf h} \uparrow \m :=
\sum_{r=1}^s
{\bf h}[r] \, \ket{j_r}\,,
\end{equation}
where ${\bf h}[r]$ is the $r$th entry of the vector ${\bf h}$,
$\{j_1,j_2,\ldots,j_s\}$ the support of $\m$ with the ordering
$j_1<j_2<\ldots<j_s$ and $\ket{j_r}$ the $j_r$th standard basis vector
of $\C^d$. A less formal way to define this vector is: the first
non-zero entry of $\m$ is replaced by the first entry of ${\bf h}$,
the second non-zero entry of $\m$ by the second entry of ${\bf h}$,
etc.

This operation is best illustrated by a simple example:
\[
\m:=
\left(
\begin{array}{c}
1 \\
0 \\
1 \\
0 \\
0 \\
0 \\
0 \\
0 \\
1 \\
\end{array}
\right)\in\{0,1\}^9\,,
\quad\quad
{\bf h}:=
\left(
\begin{array}{c}
1 \\
\omega \\
\omega^2 
\end{array}
\right)\in\C^3\,,
\quad\quad
{\bf h} \uparrow \m:=
\left(
\begin{array}{c}
1 \\
0 \\
\omega \\
0 \\
0 \\
0 \\
0 \\
0 \\
\omega^2
\end{array}
\right)\in\C^{9}
\]

\begin{theorem}[Construction of MUBs]${}$\\
Let
$\{\m_{11},\ldots,\m_{1s},\m_{21},\ldots,\m_{2s},\ldots,\m_{k1},\ldots,\m_{ks}\}$
be a $(k,s)$-net and $H$ an arbitrary generalized Hadamard matrix of
size $s$. Then the $k$ sets for $b=1,\ldots,k$
\begin{equation}\label{eq:ourMUBs}
{\cal B}_b := \left\{
\frac{1}{\sqrt{s}} 
\big( 
{\bf h}_l \uparrow \m_{bi}
\big)\,\, |\,\, l=1,\ldots,s\,,\,\, i=1,\ldots, s
\right\}
\end{equation}
are $k$ mutually orthogonal bases for the Hilbert space $\C^d$.
\end{theorem}
\proof{All vectors ${\bf h}_l \uparrow \m_{bi}$ have length $\sqrt{s}$
because all incidence vectors $\m_{bi}$ have Hamming weight
$s$. Therefore, the vectors in eq.~(\ref{eq:ourMUBs}) are normalized.

Let us now consider one fixed block $b$. Then the vectors ${\bf h}_l
\uparrow \m_{bi}$ and ${\bf h}_{l'} \uparrow \m_{bi'}$ are orthogonal
for $(l,i)\neq (l',i')$. This is seen as follows: If $i\neq i'$ then
orthogonality follows from the fact that the incidence vectors
$\m_{bi}$ and $\m_{bi'}$ have disjoint supports due to
eq.~(\ref{eq:disjoint}). If $i=i'$ and $l\neq l'$ then orthogonality
follows from the fact that ${\bf h}_l$ is orthogonal to ${\bf
h}_{l'}$. This shows that each ${\cal B}_b$ is an orthogonal basis of
$\C^d$.

Now let us consider two different blocks $b$ and $b'$. Then for
arbitrary $i,i'=1,\ldots,s$ and $l,l'=1,\ldots,s$ there is exactly one
pair $(r,r')$ such that $j_r$, $j'_{r'}$ are both $1$, where ${\rm
supp}(\m_{bi})=\{j_1,\ldots,j_s\}$ and ${\rm
supp}(\m_{b'i'})=\{j'_1,\ldots,j'_s\}$. This is ensured by
eq.~(\ref{eq:one}). Therefore, we have for the scalar product of the
embedded vectors
\[
\scalar{{\bf h}_l \uparrow \m_{bi}\,}{\,{\bf h}_{l'} \uparrow \m_{b'i'}\,} =
{\bf h}_l[r] \cdot \overline{{\bf h}_{l'} [r']}\,.
\]
It is the product of two complex numbers of modulus one. This proves
that the square norm of the scalar product of $\frac{1}{\sqrt{s}}{\bf
h}_l \uparrow \m_{bi}$ and $\frac{1}{\sqrt{s}}{\bf h}_{l'} \uparrow
\m_{b'i'}$ is $1/s^2$. Consequently, the bases ${\cal B}_b$ and ${\cal
B}_{b'}$ are mutually unbiased for $b\neq b'$.}

\begin{example}[MUBs for $d=4$]\label{ex:MUB4}${}$\\
Let us apply our construction to the incidence vectors in
Example~\ref{ex:incidence} and the Hadamard matrix of size $2$
\[
H:=
\left(
\begin{array}{rr}
1 &  1 \\
1 & -1
\end{array}
\right)
\]
Then we obtain the following three mutually unbiased bases:
\[
{\cal B}_1 := \left\{
\frac{1}{\sqrt{2}}\left(
\begin{array}{r}
1 \\
1 \\
0 \\
0 
\end{array}
\right)\,,
\frac{1}{\sqrt{2}}\left(
\begin{array}{r}
1 \\
-1 \\
0 \\
0 
\end{array}
\right)\,,
\frac{1}{\sqrt{2}}\left(
\begin{array}{r}
0 \\
0 \\
1 \\
1 
\end{array}
\right)\,,
\frac{1}{\sqrt{2}}\left(
\begin{array}{r}
0 \\
0 \\
1 \\
-1 
\end{array}
\right)
\right\}\,,
\]
\[
{\cal B}_2 := \left\{
\frac{1}{\sqrt{2}}\left(
\begin{array}{r}
1 \\
0 \\
1 \\
0 
\end{array}
\right)\,,
\frac{1}{\sqrt{2}}\left(
\begin{array}{r}
1 \\
0 \\
-1\\
0 
\end{array}
\right)\,,
\frac{1}{\sqrt{2}}\left(
\begin{array}{r}
0 \\
1 \\
0 \\
1 
\end{array}
\right)\,,
\frac{1}{\sqrt{2}}\left(
\begin{array}{r}
0 \\
1 \\
0 \\
-1 
\end{array}
\right)
\right\}\,,
\]
\[
{\cal B}_3 := \left\{
\frac{1}{\sqrt{2}}\left(
\begin{array}{r}
1 \\
0 \\
0 \\
1 
\end{array}
\right)\,,
\frac{1}{\sqrt{2}}\left(
\begin{array}{r}
1 \\
0 \\
0 \\
-1
\end{array}
\right)\,,
\frac{1}{\sqrt{2}}\left(
\begin{array}{r}
0 \\
1 \\
1 \\
0 
\end{array}
\right)\,,
\frac{1}{\sqrt{2}}\left(
\begin{array}{r}
0 \\
1 \\
-1\\
0  
\end{array}
\right)
\right\}
\]
Note that, as mentioned before, the construction is not optimal in
this case.
\end{example}

\section{Construction of the incidence vectors}\label{sec:equivalence}

The problem of finding a large collection of incidence vectors
satisfying the conditions in eq.~(\ref{eq:disjoint}) and
eq.~(\ref{eq:one}) is a well studied problem in design theory. We have
already mentioned that such a collection is equivalent to a net. There
are many more objects that are equivalent: transversal designs,
mutually orthogonal Latin squares, and orthogonal arrays
\cite{BJL:99I,BJL:99II,CD:96}. We concentrate on the equivalence of
mutually orthogonal Latin squares and nets.

\subsection*{Mutually orthogonal Latin squares}
We refer the reader to \cite{BJL:99I,BJL:99II,CD:96,HSS:99} for an
extensive introduction to Latin squares. A {\it Latin square of order
$s$} is an $s\times s$ array with entries from a set $S$ of
cardinality $s$ such that each element of $S$ appears equally often in
every row and every column. It is easily seen that a Latin square of
order $s$ exists for every positive integer $s$. For example, one may
label the rows and columns by $0,1,\ldots,s-1$ and define the entry
$L_{ij}$ in row $i$ and column $j$ to be $i+j$, where the addition is
modulo $s$. The resulting Latin square for $s=4$ is shown below:
\[
\begin{array}{cccc}
0 & 1 & 2 & 3 \\
1 & 2 & 3 & 0 \\
2 & 3 & 0 & 1 \\
3 & 0 & 1 & 2 \\
\end{array}
\]
Two Latin squares $L$ and $L'$ of order $s$ are said to be {\it
orthogonal} to each other if when one is superimposed on the other the
ordered pairs $(L_{ij},L'_{ij})$ of corresponding entries consist of
all possible $s^2$ pairs. A collection of $w$ Latin squares of order
$s$, any pair of which is orthogonal, is called a set of {\it mutually
orthogonal Latin squares} (MOLS). We use $N_{MOLS}(s)$ to denote the
maximal value $w$ such that there are $w$ MOLS of order $s$. It is
known that $N_{MOLS}(s)\le s-1$ for all $s$. If this bound is attained
we say that there is a {\it complete} set of mutually orthogonal Latin
squares of order $s$.

A construction for complete sets of MOLS of order $s$ is known if $s$
is a prime power \cite[Theorem~8.3]{HSS:99}. No construction for a
complete set of mutually orthogonal Latin squares for any other value
is known at the present time. In fact, if $s$ is not a prime power,
the largest value of $w$ for which it is known that there are $w$ MOLS
of order $s$ is usually considerably smaller than $s-1$, and for
several values it is known that $s-1$ MOLS of order $s$ cannot
exist. A table with the largest known values for $w$ is presented in
\cite{ABCD:96} for $s< 10\, 000$. In \cite{HSS:99} these numbers are
shown for $s\le 100$.

Let us now summarize some known results on the number of MOLS of a
given order. A result due to {\sc Wilson} \cite{BJL:99I} on the
existence of six MOLS shows that there are are $w\ge 6$ MOLS for
orders $s$ with $s\ge 76$. {\sc Chowla, Erd\"os and Strauss}
\cite{CES:60} proved the important fact that
\[
\lim_{s\rightarrow\infty} N_{MOLS}(s) = \infty\,.
\]
They were even able to show that there is a number $s_0$ so that we have
for all $s>s_0$
\begin{equation}\label{eq:asymptotic}
N(s)\ge \frac{1}{3} s^{1/91}
\end{equation}
{\sc Wilson} \cite{Wilson:74} improved this to $N_{MUB}(s)\ge
s^{1/17}$, and {\sc Beth} \cite{Beth:83} obtained the exponent
$\frac{1}{14.8}$. These proofs involve number-theoretic sieve methods.

We have the following important equivalence (see \cite[Remark~2.10 and
Theorem~2.12]{CD:96}:

\begin{fact}
The existence of $w$ mutually orthogonal Latin squares is equivalent
to the existence of an $(s,k)$-net with $k=w+2$.
\end{fact}

\section{How good is the construction?}
Example~\ref{ex:MUB4} shows that our construction yields maximally
three mutually orthogonal bases for dimension $d=2^2$ because there
are no $w$ MOLS of order $2$ with $w>1$. The maximal value
$N_{MUB}(4)$ is $5$ because $4$ is a prime power (use the optimal
construction in \cite{Ivanovic:81,WF:89,BBRV:02} for prime powers).

More generally, it is clear that our construction can never beat the
``prime power'' construction for dimension of the form $d=p^{2e}$,
where $p$ is an arbitrary prime and $e\ge 1$. Our construction gives
$p^e+1$ MUBs because there are $p^e-1$ MOLS of order $p^e$, whereas
the prime power construction yields $p^{2e}+1$ MUBs.

The advantage of our construction is that it yields more MUBs for many
square dimensions than the ``reduce to prime power'' construction. Let
us characterize the cases in which it is better.

Let $d=s^2$ be a square dimension, i.e., its prime power factorization
has the form
\[
d=p_1^{2 e_1} p_2^{2 e_2} \cdots p_r^{2 e_r}\,,
\]
where the $p_i$'s are different primes and $e_i>0$ for
$i=1\,\ldots,r$. Then the ``reduce to prime powers'' gives $n+1$ MUBs,
where
\begin{equation}\label{eq:minimalPP}
n:=\min_{i=1,\ldots,r} \{p_i^{2 e_i}\}\,.
\end{equation}
To see if our construction yields more MUBs for the dimension $d=s^2$
we have to check if
\begin{equation}\label{eq:necessaryIneq}
N_{MOLS}(s)+2>n+1\,
\end{equation}
where $n$ is the minimal prime power contained in $s^2$ (as given by
eq.~(\ref{eq:minimalPP})). The first $s$ for which we know that this
is true is $s=26$. We have \cite[Theorem~2.44 in Chapter II]{CD:96}

\begin{fact}
$N_{MOLS}(26)\ge 4$. 
\end{fact}
Therefore, we can construct $N_{MOLS}(26)+2\ge 6$ MUBs in dimension
$d=26^2$, whereas the ``reduce to prime power'' construction gives
only $5$ MUBs since $4$ is the smallest prime power contained in $d$.

More generally, let us consider the numbers 
\[
{\cal M}:= \{s\in{\mathbb N}\,\, | \,\, s \equiv 2 \mbox{ mod } 4\}\,.
\]
For each $s\in {\cal M}$ the minimal prime power contained in $s$ is
$2$. Therefore, the ``reduce to prime powers'' construction gives only
$5$ MUBS in dimension $d=s^2$. Due to Wilson's result we know that
$N_{MOLS}(s)\ge 6$ for all $s\ge 76$. Therefore, our construction
gives at least $8$ MUBs for dimensions $d=s^2$ for $s\ge 76$. This
shows that there are infinitely many dimension in which our
construction yields better results. 

It follows from our construction and the asymptotic result on the growth of
mutually orthogonal Latin square that:
\begin{corollary}
We have $N_{MUB}(s^2)\ge s^{1/14.8}$ for all $s$ but finitely many
exceptions.
\end{corollary}
To determine the cases in which our construction is better one should
consult Table~2.72 in \cite[Chapter II]{CD:96}. This table shows the
best known lower bounds on $N_{MOLS}(s)$ for all $s\le 10\, 000$.

It is important that our construction can also be combined with the
idea of the ``reduce to prime power'' construction. Instead of
decomposing the dimension $d$ into prime powers we may decompose it
into squares and prime powers. This could be advantageous e.g.\ for
numbers of the form $d=s^2 p$, where $s^2$ contains a small prime
power and $p$ is a prime number that is relatively prime to $s^2$ and
is larger than the best known lower bound on $N_{MUB}(s)$. Consider
e.g.\ the dimension $d=26^2 \cdot 7$. Then the ``reduce to prime
powers'' construction gives only $5$ MUBs. We can construct $6$ MUBs
in dimension $26^2$ with our construction and $8$ MUBs in dimension
$7$. Therefore, we obtain $6$ MUBs in dimension $d=26^2 \cdot 7$.
This shows that our construction has implications not only for square
dimensions.

\subsubsection*{Acknowledgments}
We would like to thank Markus Grassl, Betina Schnepf and Thomas Decker
for helpful discussions.

\end{document}